\documentclass{nature}
\usepackage{graphicx} 
\usepackage{cite}
\usepackage{epsfig} 
\usepackage{amsmath}
\usepackage{xcolor}
\usepackage{soul}
\usepackage{ulem}
\usepackage{caption}
\usepackage{xr}

\title{Active beam steering enabled by photonic crystal surface emitting laser}

\author{Mingjin Wang$^{1,5,8}$, Zihao Chen$^{2,8}$, Yuanbo Xu$^1$, Jingxuan Chen$^1$, Jiahao Si$^1$, Zheng Zhang$^6$, Chao Peng$^{2,7*}$ \& Wanhua Zheng$^{1,3,4,5*}$}
\begin{document}
\maketitle

\begin{affiliations}
\item Laboratory of Solid State Optoelectronics Information Technology, Institute of Semiconductors, CAS, Beijing 100083, China
\item State Key Laboratory of Advanced Optical Communication Systems and Networks, Department of Electronics \& Frontiers Science Center for Nano-optoelectronics, Peking University, Beijing 100871, China
\item College of Future Technology, University of Chinese Academy of Sciences, Beijing 101408, China
\item Center of Materials Science and Optoelectronics Engineering, University of Chinese Academy of Sciences, Beijing 100049, China
\item State Key Laboratory on Integrated Optoelectronics, Institute of Semiconductors, CAS, Beijing 100083, China 
\item Huawei Technologies Co. Ltd., Wuhan 430070, China
\item Peng Cheng Laboratory, Shenzhen 518055, China
\item These authors contributed equally to this work

\end{affiliations}
Corresponding Author: Chao Peng (email: pengchao@pku.edu.cn) \& Wanhua Zheng (email: whzheng@semi.ac.cn)

\begin{abstract}
Emitting light towards on-demand directions is important for various optoelectronic applications, such as optical communication, displaying, and ranging. However, almost all existing directional emitters are assemblies of passive optical antennae and external light sources, which are usually bulky, fragile, and with unendurable loss of light power. Here we theoretically propose and experimentally demonstrate a new conceptual design of directional emitter, by using a single surface-emitting laser source itself to achieve dynamically controlled beam steering. The laser is built on photonic crystals that operates near the band edges in the continuum. By shrinking laser sizes into tens-of-wavelength, the optical modes quantize in three-dimensional momentum space, and each of them directionally radiates towards the far-field. Further utilizing the luminescence spectrum shifting effect under current injection, we consecutively select a sequence of modes into lasing action and show the laser maintaining in single mode operation with linewidths at a minimum of $1.8$ MHz and emitting power of $\sim$ ten milliwatts, and we demonstrate fast beam steering across a range of $3.2^\circ \times 4^\circ$ in a time scale of $500$ nanoseconds. Our work proposes a novel method for on-chip active beam steering, which could pave the way for the development of automotive, industrial, and robotic applications.

\end{abstract}

\section{Introduction}
Optical beam steering is an eagerly desired functionality for various optoelectronic applications, such as optical switching \cite{andersen_beam_2022, smalyukh_three-dimensional_2010}, displaying \cite{wu_dynamic_2019, zheng_three-dimensional_2016, zhu_flexible_2015}, printing \cite{braun_photochemistry_2021}, and light detection and ranging (Lidar) \cite{lum_ultrafast_2020, jiang_time-stretch_2020, li_single-photon_2021, whitworth_solution-processed_2021}. To date, almost all existing methods of beam steering are built upon passive systems in which the light is supplied by external sources. Examples include applying micro-electro-mechanical system \cite{zhang_large-scale_2022, lihachev_low-noise_2022}, liquid crystal \cite{li_phase-only_2019, park_all-solid-state_2021}, silicon photonics \cite{yang_inverse-designed_2020, zhang_large-scale_2022, xiang_high-performance_2021, guidry_quantum_2022, liu_photonic_2020}, or true time delay \cite{lee_ultra-low-loss_2012, jiang_novel_2017} to realize optical phased array \cite{rogers_universal_2021, sun_large-scale_2013} and focal-plane switch array architectures \cite{fu_light_2022}. Besides numerous advances, one prominent limitation of passive beam steering systems stems from their unendurable loss of light power, which consequently hinders the promotion of detection sensitivity and ranges for practical applications. Inspired by active electronically scanned array radars in radio-frequency (RF) spectrum \cite{friis_multiple_1937}, active beam steering enabled by semiconductor light emitters becomes a promising approach, owing to their particular advantages in high power emission and potentials of on-chip integration. A pioneering work had conceptually demonstrated directional beam emissions of lasers by using composite photonic-crystal (PhC) structures \cite{kurosaka_-chip_2010}, and further, dynamical beam scanning has been achieved by assembling many lasers and circuit-driving them in a matrix configuration \cite{sakata_dually_2020}. However, to the best of our knowledge, using a single laser to realize actively controlled beam steering yet remains a challenge. 

Here we theoretically propose and experimentally demonstrate a method of active beam steering by using a single photonic-crystal surface-emitting laser (PCSEL). Specifically, we design a PhC laser operating in radiation continuum, whose sizes are in tens-of-wavelengths scale along both transverse and vertical directions. Accordingly, the continuous bulk band of PhC turns into a set of discrete optical modes that possess quantized momentum in three dimensions and emit directionally. When multiple quantum wells (MQWs) as optical gain materials are injected by different currents, the peak of gain spectrum shifts, and thus the device selects one of the quantized modes for each current into lasing oscillation and radiating towards a given off-normal direction. By dynamically controlling such an operation, a sequence of modes consecutively lase in a real-time manner, realizing fast active beam scanning and steering in a time scale of $500$ nanoseconds.

\section{Design and principles}
\setcounter{equation}{0}
\renewcommand{\theequation}{\arabic{equation}}

\begin{figure}
\centering
\includegraphics[width=16cm]{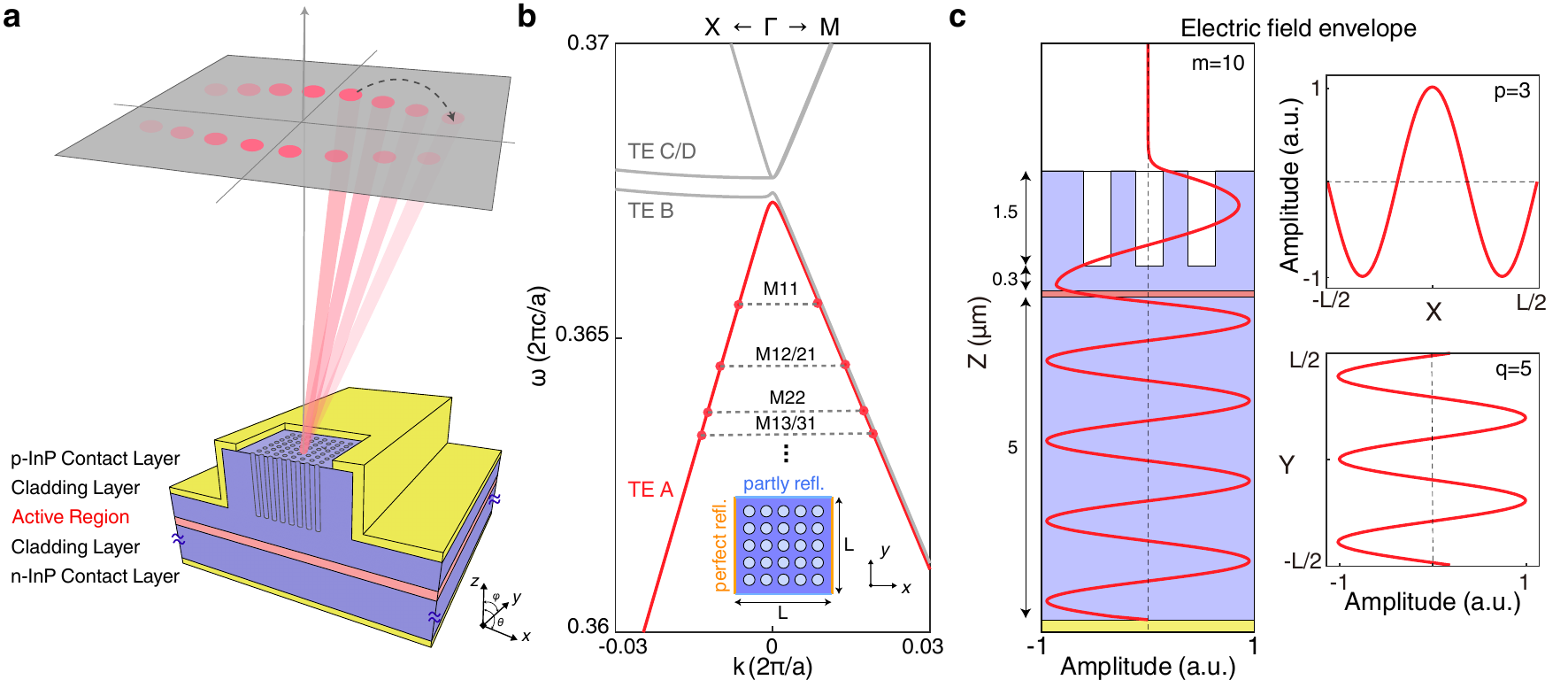}
\caption{\textbf{Mode quantization in PCSEL for beam steering}.
\label{mfig1}
(a) 
Schematic of the designed PCSEL, which is electrically pumped to produce directional spot-like patterns in the far-field.
(b)
Band structure of the PhC, in which four bands TE-A to D reside at the band edge in the continuum near the 2nd-order $\Gamma$ point. The inset shows the real space geometry of finite size PhC, with perfect and partly reflective boundaries in $x$ and $y$ directions, respectively. Accordingly, the bands are quantized to discrete modes, denoted as M$(p,q)$.
(c)
The electric field envelopes in vertical and transverse directions, calculated by coupled-wave-theory (CWT). The envelopes fold in $z$ direction (left), as well as  $x$ and $y$ directions (right), depicted by quantum numbers $(m,p,q)$, respectively.
}
\end{figure}

We design a PCSEL structure as schematically shown in Fig.~\ref{mfig1}a, in which circular air-holes of square lattice are patterned in the center of a ridge region. The PhC structure locates on the top of InP epi-growth wafer, and AlGaInAs MQWs layers are sandwiched in the center of wafer as gain medium. The top of ridge region and bottom of substrate are covered with metals for current injection, leaving a window on the top of PhC for light emission. Comparing with other reported PCSELs \cite{hirose_watt-class_2014,yoshida_double-lattice_2019,shao_high-performance_2020,mao_magic-angle_2021,imada_coherent_1999,grim_continuous-wave_2014,ellis_ultralow-threshold_2011,matsubara_gan_2008,colombelli_quantum_2003,ha_directional_2018,song_single-fundamental-mode_2002}, our air-holes locate on the top of the wafer rather than embedded inside, and they are in relatively high aspect-ratio (depth $h=1580$ nm and radius $r=167$ nm). In our design, the PhC structures consist of $84\times 84$ air-holes at a periodicity of $a=540$ nm. We assume the boundaries in $x$ direction are perfectly reflective because of metal coating, but the others are partially reflective on the dielectric ridge facets in $y$ direction (see the Supplementary Section I for the details of design). Under lasing oscillation, light near-normally emits from the surface of PhC, and generates a multi-spot beam pattern in the far-field (gray plane, Fig.~\ref{mfig1}a). 

We calculate the bulk bands of PhC by assuming an infinite periodic unit-cell (COMSOL Multiphysics), as shown in Fig.~\ref{mfig1}b. The laser operates near the band-edge at the second-order $\Gamma$ point, where there are four bands denoted as TE-A to D. Among these bands, TE-A and B are in high quality factors ($Q$s) owing to their anti-symmetric nature, while right at the $\Gamma$ point, they are recognized as symmetry-protected bound states in the continuum (BICs) \cite{hsu_observation_2013,jin_topologically_2019,hsu_bound_2016}. In our design, TE-A band is expected to lase (red line, Fig.~1b). Because the boundaries are reflective in lateral directions, the light is transversely confined into a small volume, and thus the continuous TE-A band quantizes into discrete modes with a mode spacing of $\pi k = \pi/L$ , where $L$ is the length of PhC region \cite{chen_observation_2022, chen_analytical_2022}. Consequently, each mode can be labeled by a pair of integers $(p,q)$ (dashed lines, Fig.~\ref{mfig1}b), indicating that its momentum is mostly localized near ($p\pi/L, q\pi/L$) in the first quadrant of the momentum space. We further calculate the envelops of electric fields in both vertical and transverse directions (Fig.~\ref{mfig1}c) by using a three-dimensional (3D) coupled-wave theory (CWT) method \cite{liang_three-dimensional_2011,peng_three-dimensional_2012,yin_analytical_2017} (see the Supplementary Section II for the details), in which we conceptually apply a thinner substrate of $5~\mu$m instead of $135~\mu$m in our realistic device for simplicity. Thanks to the metal mirror at the bottom, such a vertical stack is still capable of confining light along the out-of-plane direction (left panel, Fig.~1c), even in the case of the low refractive-index PhC layer residing at the most top. It is noticed that the envelop in $z$ direction folds many times according to the distance between the top surface and bottom mirror, which in fact rises a bunch of guide-modes identified by a quantum number of $m$ in $z$ direction, depicting the number of maximum on the envelop. The theory indicates that (Supplementary Section II), each of guided-modes corresponds to a set of bulk bands that are featured in transverse quantization. Therefore we conclude that the optical modes in our design are discrete in all the three dimensions, represented by a triple of integers $(m,p,q)$.

\begin{figure}
\centering
\includegraphics[width=16cm]{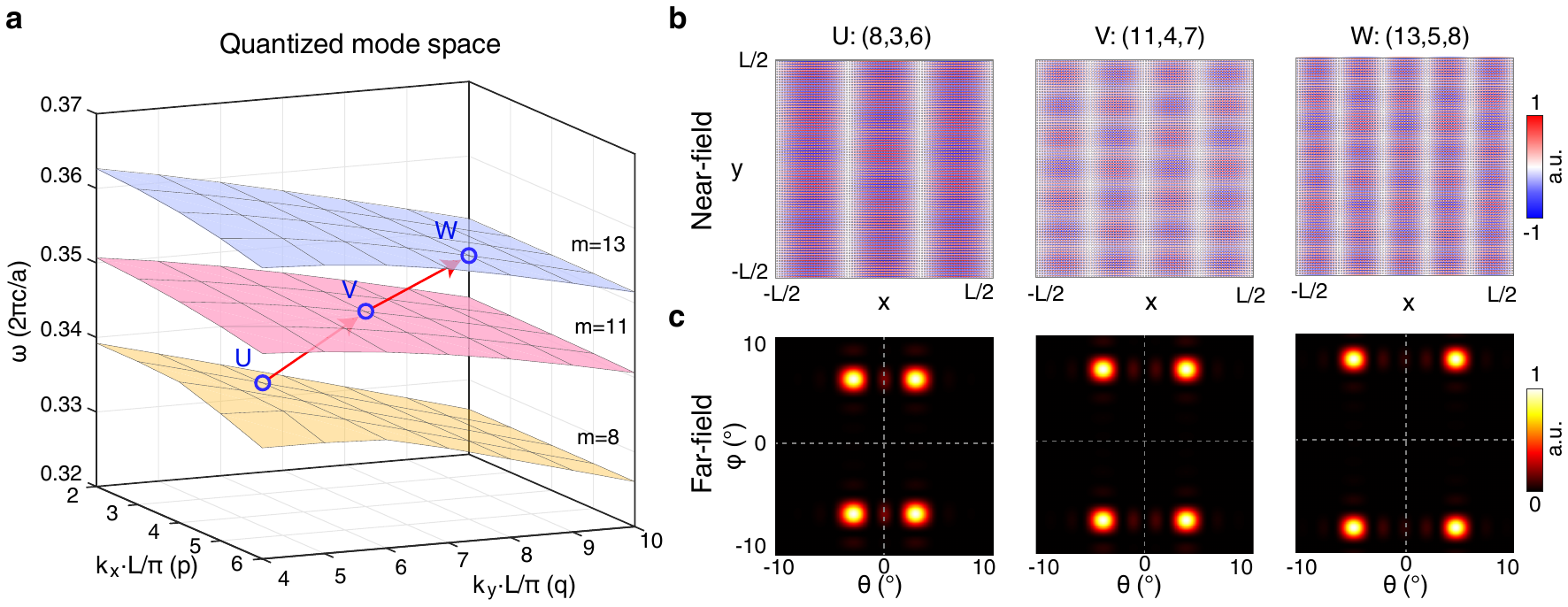}
\caption{ \textbf{Mode switching in the designed PCSEL}
\label{mfig2}
(a) A zoom-in-view of band TE-A near the $\Gamma$ point, where the continuous band is quantized to discrete optical modes in 3D momentum space, labeled by a triple of integers $(m,p,q)$. The red arrows show a sequence of mode switching from U, to V, then to W, indicating that the quantum numbers change from $(8,3,6)$, to $(11,4,7)$, then to $(13,5,8)$, correspondingly. 
(b)
The near-field patterns of modes U, V, W in the finite-size PhC region. The nodal lines show the modes are quantized in the transverse directions
(c)
The far-field patterns of modes U, V, W in terms of off-axis angles $(\theta,\phi)$, showing directional-emitting characteristics of the PCSEL. All the results are calculated by using CWT.
}
\end{figure}

To illustrate the details of mode quantization, we present  a zoom-in-view of band structure in the vicinity of $\Gamma$ point (Fig.~\ref{mfig2}a), in which the in-plane wavevectors are normalized to the lateral size $L$ to equivalently depict the quantum numbers $(p,q)$ in $x$ and $y$ directions, respectively. Three surfaces (yellow, red and blue) represent the sub-bands spawn from TE-A band but have their own quantum numbers $m$ in $z$ direction. Each of sub-bands is further quantized by in-plane confinement, as shown by the gray grids. We take three discrete modes U, V, and W, for examples, as well as plot their near- and far-field patterns that are calculated from 3D CWT (Fig.~\ref{mfig2}b and c). Clearly, the near-fields of modes exhibit multiple nodal lines, and their corresponding far-field patterns show highly-directional characteristics towards different off-normal angles $\theta$ and $\phi$ along $x$ and $y$ directions, governed by their unique quantum numbers $(m,p,q)$, respectively. As a result, by controlling the device to consecutively lase in a sequence from modes U to W, we are able to demonstrate active beam steering action in the far-field. Note that each discrete mode possesses a given eigen-frequency owing to band dispersion, so we propose that utilizing the gain-peak shifting effect \cite{pauzauskie_semiconductor_2006} under different carrier densities would be a feasible way to dynamically select one mode at a time for lasing. Such a way innovates a method of beam steering, which would be experimentally demonstrated in the following section. 

\section{Sample fabrication and experimental characterization}
To verify the design and principles, we fabricate the PCSEL samples on multi-layered epi-wafer of InP in which an AlGaInAs MQWs layer of $\sim90$ nm serving as a gain-medium. The PhC structures are patterned on a ridge region in a width of  $90~\mu$m  and a height of $1.8~\mu$m, with etching depth of around $1.58~\mu$m through the top, by using e-beam lithography, dry etch process, plasma-enhanced chemical vapor deposition silicon dioxides growth, metal deposition, then followed by lift-off (see Methods for details). The substrate thickness is about  $135~\mu$m with its bottom side coated by metal materials. Comparing to other realization of PCSELs \cite{hirose_watt-class_2014,yoshida_double-lattice_2019}, our design doesn't require sophisticated re-growth process to bury air-holes into the middle of the device, and thus significantly simplifies the fabrication process. The PhC periodicity of $a=540$ nm and radius of $r=167$ nm result in our lasers operating at telecommunication and eye-safe wavelength of around $1550$ nm. A top-view of PCSEL sample is observed by using an optical microscope, showing a part of the ridge region and the whole PhC region in size of $45 \times 45 ~\mu$m$^2$ (Fig.~\ref{mfig3}a). The scanning electron microscope image of PhC pattern is presented in Fig.~\ref{mfig3}b, in which a few air-holes are cleaved by focus ion beam for better side-viewing (see Fig.~\ref{mfig3}c).

We first evaluate the feasibility of shifting the gain peak for MQWs via changing the injection current. In theory, the Fermi level of semiconductor would be lifted by carrier densities, thus modifying the center of gain in luminescence spectrum \cite{sze_physics_2006, coldren_diode_2012}. In MQWs structure, considering the band-gap shrinkage effect, the increasing injection current with a higher carrier density leads to a rigid shift of the entire gain spectrum to a longer wavelength. To characterize such an effect, we fabricate a bared ridge device by coating the same metal contacts as the PCSEL device but without etching the PhC patterns in the center of the opening window. As shown in Fig.~\ref{mfig3}d, the gain spectra under the injection currents exhibit a typical full width at half maximum (FWHM) of $\sim 12$ nm. Their peaks red-shift in a range of $\sim 35.6$ nm when the current increases from $350$ mA to $980$ mA, and thus clearly proves the validity of the gain-peak-shifting effect. 

Further, we characterize the lasing behaviors of the PCSEL sample. By increasing the current, a lasing threshold has been reached at $180$ mA, identified by the transition point of slope on the Current-Power (I-O) curve (Fig.~\ref{mfig3}e). Due to the high density of states (DOS) originated from mode quantization, the passive $Q$s of discrete optical modes are hard to be measured from the spectrum at lasing threshold. Nevertheless, we believe that the device behaves as our theoretical predication in $Q\sim1000$, estimated from its relatively high value of lasing threshold \cite{liu_high-q_2019, liang_three-dimensional_2012, chua_low-threshold_2011, chen_observation_2022} (see Supplementary Section II). By further increasing the current, a slope efficient of $\sim15.75$ mW/A has been extracted from the I-O curve, with an output power of $12.6$ mW at $980$ mA. At the highest injection current, we find the laser device still operates in single-mode oscillation, supported by the delayed self-heterodyne measurement of linewidth (see Method for details). As illustrated in Fig.~\ref{mfig3}f, the linewidth of $1.8$ MHz is achieved under the current of $980$ mA. At the same current, we record the emission pattern of the PCSEL by directly imaging the far-field via a relay 4$f$ optical system (see Method for details), as shown in Fig.~\ref{mfig3}g. Four bright and distinct light spots are observed at off-normal wavevectors in the field-of-view, proving that the laser indeed operates in single-mode and emits directionally as the prediction of our theory. Importantly, during increasing injection current, we find that the lasing wavelength remains as a single peak for each current but red-shifts correspondingly (in Fig.~\ref{mfig3}h), following the same trend of gain-peaks-shifting in the bared ridge device (Fig.~\ref{mfig3}d). These experimental results indicate that our PCSEL device is truly consecutively lasing in a sequence of discrete modes across the quantized mode space as designed. More analysis and discussion are presented in Supplementary Section III.

By dynamically driving the laser by different currents, we demonstrate that the emission beams are real-timely controlled to scan in the far-field. Given the emission patterns are mirror-symmetric in $x$ and $y$ directions, we only plot the first quadrants of the far-fields during the scanning, as shown in Fig.~\ref{mfig4}a, in which several beams are overlapped and marked by circles. By increasing the current from $350$ mA to $980$ mA, each of the lasing modes in Fig.~\ref{mfig3}h remains as a single spot in the first quadrant at its corresponding current with a typical diverge-angel of around $0.68^\circ$, the wavelength red-shifts from $1557.9$ nm to $1594.6$ nm (see Fig.~\ref{mfig4}b), and the emission angles $(\theta,\phi)$ semi-continuously move from $(2.9^\circ, 18.6^\circ)$ to $(6.1^\circ, 14.6^\circ)$ in the far-field (see Fig.~\ref{mfig4}c), showing a 2D scanning range of  $3.2^\circ \times 4^\circ$ that are calibrated by the numerical aperture (N.A) of the object lens (see Supplementary Video for the details). Further, we characterize the quantum numbers $(m,p,q)$ of modes in the quantized mode space from fitting the measured emission angles and the lasing wavelengths by matching the calculations based on our theory, as illustrated in Fig.~\ref{mfig4}d. As expected, during increasing the current, the lasing modes successively jump in a zigzag-like sequence of quantized momenta to the lower values and their corresponding lasing wavelengths monotonically red-shift to the higher values. Because the substrate with a coating metal mirror is considerably thick, the modes own large quantum numbers $m\sim 320$ in $z$ direction. But much smaller quantum numbers $p\sim 4$ and $q\sim 16$ are found since the device sizes are relatively compact in the transverse directions. Noteworthy that, thanks to the unique quantization in $z$ direction which is different from those conventional PCSELs, the momentum increment between two adjacent modes is quite small, namely the semi-continuous measured typical angle-steps of $0.3^\circ$ in $x$ direction and $0.8^\circ$ in $y$ direction as the beam scanning.

At last, we experimentally evaluate the speed of beam steering. According to the theory, the spot-like emission patterns are given by discrete modes with quantized momenta as $(m,p,q)$. Therefore, the beam scanning behavior actually corresponds to the transition between the modes with different quantum numbers. Such a transition doesn't require an abrupt change of electromagnetic field in a unit-cell (for instance from mode A to B), but merely a re-distribution of their slow-varying envelops of the light and carriers. Similar phenomena had been demonstrated in a ultra-fast control of  vortex microlaser, in which the switching time is about 1.5 picosecond owing to the excellent mobility of carriers in perovskite material \cite{huang_ultrafast_2020}. Therefore, we also expect our laser can switch rapidly for beam steering. 

The measurement setup is shown in Fig.~\ref{mfig5}a, in which we use two photo-diodes (PDs) to detect the beam spots in the first quadrants of the field-of-views that are given by two different modes $(343,3,19)$ and $(297,6,14)$. The two modes are spatially filtered out  by using two pin-hole diaphragms, respectively. To demonstrate fast beam steering, the laser is driven by a commercial current source with a maximum bandwidth of $1$ MHz, as the circuit model schematically shown in Fig.~\ref{mfig5}b. We apply an electrical bipolar waveform of 1 MHz to rapidly switch the injection currents between $350$ mA and $980$ mA (Fig.~\ref{mfig5}c). Owing to the bandwidth limit of the current source,  the voltages measured between the laser contacts show a sinusoidal shape with a peak-to-peak value of $1100$ mV (Fig.~\ref{mfig5}d), which agree with the static measurement of laser volt-current (V-I) curve (Supplementary Section III). From the measurements of PDs, we find the peak light intensities in turns reach their maxima in a time scale of $\sim 500~$ns (Fig.~\ref{mfig5}e), showing that the beams can scan fast under a current modulation.

\section{Discussions}
Above theory and experiments demonstrate a new method of active beam steering from selectively exciting a series of quantized modes to lasing oscillation. Compared with existing methods, our proposal has particular advantages in its high emission power (up to $\sim$ ten mWs) and compactness. Besides, we believe such a method also has much potential in further promoting the performances of divergence angle, scanning step, range and speed, in particular for Lidar applications, as we elaborate as following. 

As a conceptual realization, we apply a relatively small footprint of $45 \times 45 ~\mu$m$^2$, resulting in a typical diverge angle of  $0.68^\circ$ as well as relatively sparse scanning steps of $0.3^\circ$ in $x$ direction and $0.8^\circ$ in $y$ direction. By further expanding the footprint, higher emission power, smaller diverge angle, and finer scanning step are expected at the same time, but at a cost of a more sophisticated control of optical gain to precisely pick up the specific mode for lasing oscillation. For instance, we can design a set of finger-like electrical contacts with selectively turn-on functionality \cite{xu_high-performance_2020}, and thus the in-plane quantum number can be independently controlled for better controllability in addition to the gain-peak-shifting effect. On the other hand, the scanning range of our laser is mainly determined by the band dispersion of PhC, which can be engineered to be flatter (for instance, applying twisted PhC structures \cite{lou_theory_2021, tang_modeling_2021}) for achieving a wider range of beam steering under the same gain-peak-shifting. Noteworthy that the metal contacts, ridge geometry and MQWs layers in our laser haven't been optimized in purpose for high speed operation. We believe that our method still has much potential towards a higher operation speed, given that the mobility of electrons in MQWs is considerably high according to report results  \cite{praseuth_growth_1988} ($\sim 5,000 $ cm$^2$/Vs depending on the barrier thickness and the sample temperature).

\section{Conclusion}
To summarize, we present a new method of active beam steering by using a single photonic-crystal surface-emitting laser, in which a sequence of modes are selectively excited by gain-peak-shifting effect in a real-time manner for single-mode lasing and directional emitting. We experimentally demonstrate that the laser shows a linewidth in megahertz scale and emitting power of $\sim$ ten milliwatts, as well as supports a beam scanning range of $3.2^\circ \times 4^\circ$ in a switching time of $500~$ns under real-timely driving. Such a conceptual design reveals the new possibility of realizing dynamically controlled directional emitters without an external light source, and thus paves the way to a range of applications including ultra-compact laser displays \cite{xie_metasurface-integrated_2020}, optical multiplexing \& switching \cite{puttnam_space-division_2021, lingaraju_adaptive_2021}, and laser-radar sensing systems \cite{wu_monolayer_2015, ellis_ultralow-threshold_2011}.

\section{Methods}
\subsection{Fabrication process}
The photonic crystal is designed on an InP-based epi-wafer. The epi-wafers are grown on $n$-type InP ($001$) substrates by in-house metalorganic chemical vapor deposition (MOCVD). The schematic of epi-wafer is illustrated in SFig.~1. After $n$-type buffer InP epilayer being deposited on the substrate, a grading composition AlGaInAs epilayer with heavily Si dopants is grown, followed by the MQWs on the $n$-type graded-index separate-confinement hetero-structure (GRIN-SCH) layer. The undoped MQWs are grown at around 660°C, which consist of five layers of 6 nm compressively strained wells, and six layers of 10 nm tensile strained barriers. Then $p$-type AlGaInAs GRIN-SCH is grown on the top of the active region. Prior to the deposition of Ohmic contact layer, the PhC layer is grown. Silane and diethylzinc are used as $n$-type and $p$-type doping sources, respectively.

The epiwafers are processed into 90 $\mu$m broad ridge waveguides for the later fabrication of surface emitting laser. The waveguide height is 1.8 $\mu$m. The PhC region is about  $45 \times 45 ~\mu$m$^2$, locating at the center of the ridge waveguide. The deposited metals (TiPtAu) above the PhC region are lifted off to prevent light absorbing and reflecting. The square lattice PhC structures are patterned by e-beam lithography (EBL) and inductively coupled plasma (ICP) dry etching without regrowth. The PhC structures are exposed using JEOL JBX-8100 EBL and ZEP520A photoresist. The etching depth of PhC is optimized to be around $1.58~\mu$m. The aspect ratio of PhC is about $4.5$. During the dry etching for PhC structures, the pressure, bias power, and degree of vacuum are optimized to enhance etching selectivity for reducing isotropic etching in the holes of the PhC. The fabricated wafers are cleaved into bars with a cavity length of $500~\mu$m. The bars are coated by high-reflection facets on both edge sides, and the high-reflection coatings consist of periodic silicon oxides and titanium dioxides.

\subsection{Delayed self-heterodyne measurement of linewidth}

The spectral linewidth of the laser is measured by a standard delayed self-heterodyne interferometer technique, as shown in SFig.~4. In our experimental setup, a designed wedge-shaped lensed fiber (WF) with $6^\circ$ bevel angle and an isolator with distinct ratio of $60$ dB (IOT-H-1550A) are employed to minimize optical reflection towards the laser. The collected light in the fiber is split by a  $50:50$ fiber optical coupler (OC, TW1550R5A1-1). One beam passes through an acousto-optic modulator (AOM, SGTF80-1550-1) driven by an acousto-optic driver (SGY80-33-N-1), and the other beam passes through a single-mode (SM) fiber delay line. The fiber length is $20$ km-long ($\sim10$ kHz of measurement resolution), and AOM operates at  a constant frequency shift ($80$ MHz). The two beams are combined through the second OC, and the signal is detected by a PD (New Focus, Model 1554-B), which performs optical to electrical conversion to deliver the electrical signals to an electric spectrum analyzer (ESA, Keysight Model N9320B spectrum analyzer) through a microwave cable (MC).

\subsection{Relay 4$f$ optical system and the evaluation of the speed of beam steering}

The experimental setup is shown in SFig.~5. A laser diode controller (Stanford Research Systems, Model LDC502) connected with a function generator (Tektronix, Model AFG3252C) in a series circuit is applied to electrically drive the PCSEL through a twisted pair channel (Newport, CC306S). The lasing beam is collimated by an objective (Mitutoyo M Plan Apo NIR, 50X). After passing through a in-house relay 4$f$ system (focal lengths of $150$ mm and $200$ mm), the light is split by a beam splitter. The reflected part and the transmitted part of light are filtered separately by diaphragm $1$ and diaphragm $2$, then the splitting beams are collected by photodiode $1$ and photodiode $2$ (Thorlabs, PDA10DT-EC) in a separate manner. Diaphragm $1$ and $2$ are fixed to pick up the light spots of mode $(343,3,19)$ and $(297,6,14)$ in the first quadrants, respectively. The time-domain intensity response of two photodiodes is recorded by a mixed signal oscilloscope (Keysight, InfiniiVision MSOX6004A).

\begin{figure}
\centering
\includegraphics[width=16cm]{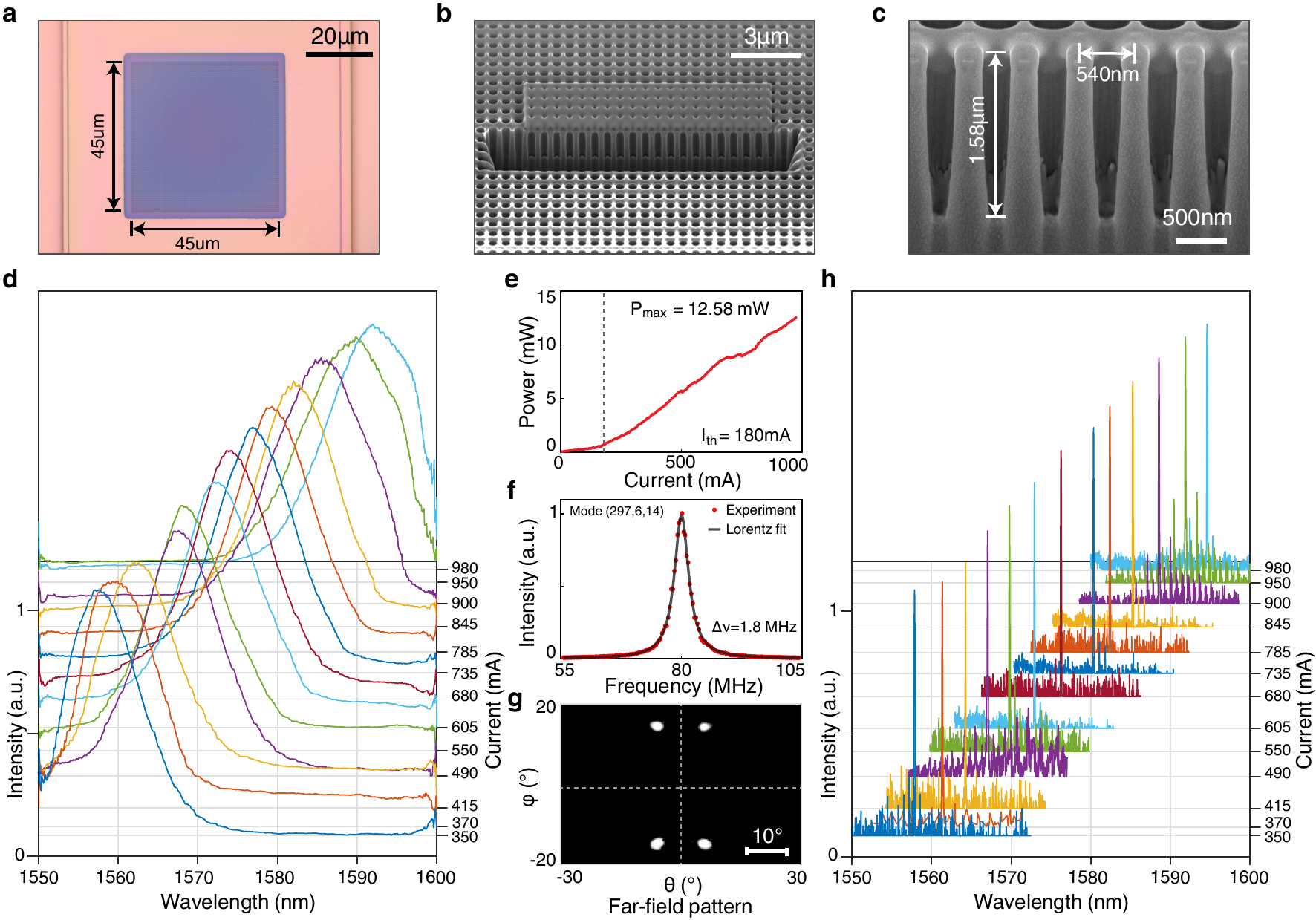}
\caption{ \textbf{Fabricated sample and lasing behaviors}. 
\label{mfig3}
(a-c) 
The optical and scanning electron microscope images of the fabricated sample with PhC structures of $a=540$ nm and $r=167$ nm.
(d)
The shifting of gain peaks in luminescence spectra when the injection currents increase from $350$ to $980$ mA. 
(e)
The measured I-O curve of the fabricated PCSEL, showing the lasing threshold at $180$ mA.
(f)
The representative linewidth of lasing mode. The RF spectrum of the beat note signal is measured at the current of $980$ mA, with resolution bandwidth (RBW) $=300$ kHz and video bandwidth (VBW)$=30$ kHz.
(g)
The far-field pattern of mode $(297,6,14)$ observed at $980$ mA. The directional angles $\theta$ and $\phi$ along $x$ and $y$ directions are  $(6.1^\circ, 14.6^\circ)$, respectively.
(h)
The evolution of lasing wavelengths that red-shift to longer wavelengths during the increasing of currents, following the same trend as luminescence spectral evolution.
}
\end{figure}

\begin{figure}
\centering
\includegraphics[width=16cm]{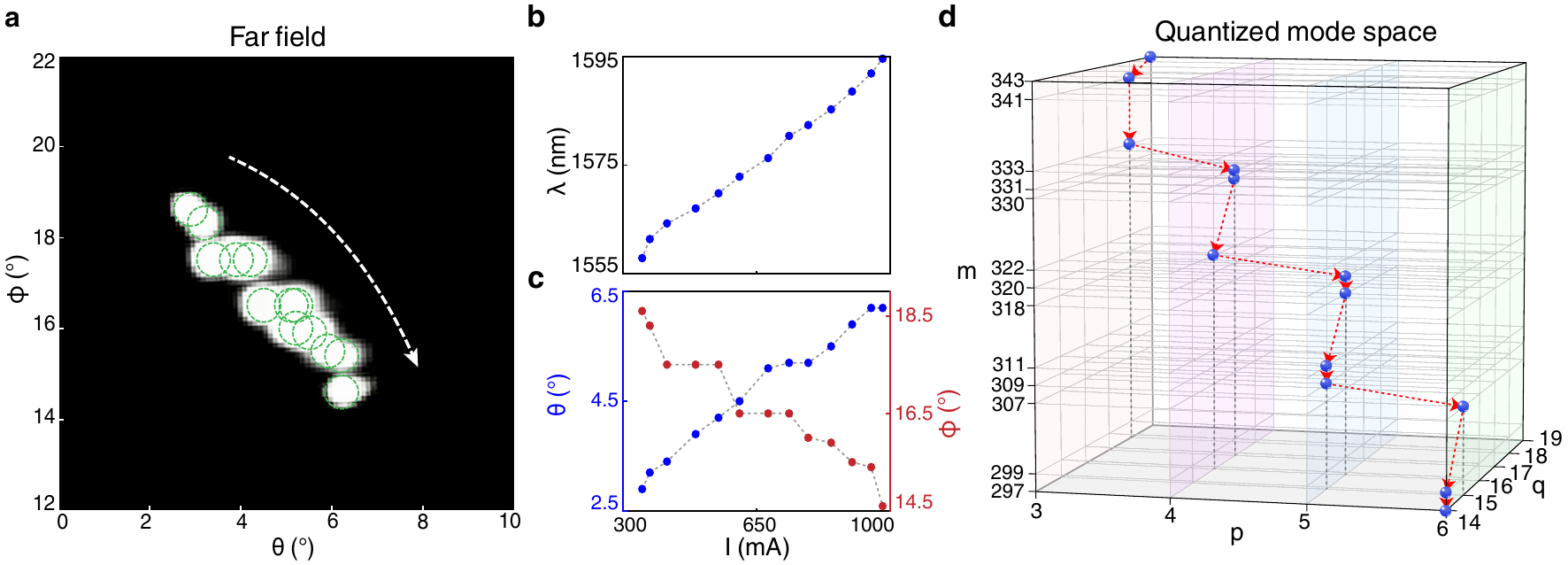}
\caption{\textbf{Observation of active beam steering of the PCSEL}
\label{mfig4}
(a) The first quadrants of the far-field, illustrating the beam steering behavior of PCSEL when the injection currents increase from $350$ to $980$ mA (guided by dashed arrow line). The dashed circles mark the beam spots at each of the currents. 
(b-c)
The peak wavelengths (royalblue) and off-normal angles in $\theta$ (blue) and $\phi$ (red) directions for the same PCSEL sample with different injection currents, showing a smooth and semi-continuous beam steering behavior of PCSEL.
(d)
The quantum numbers $(m,p,q)$ of the lasing modes in 3D momentum space based on the CWT model fitted with the experimental results, showing that the lasing modes jump in successive sequence following a zigzag route.
}
\end{figure}

\begin{figure}
\centering
\includegraphics[width=16cm]{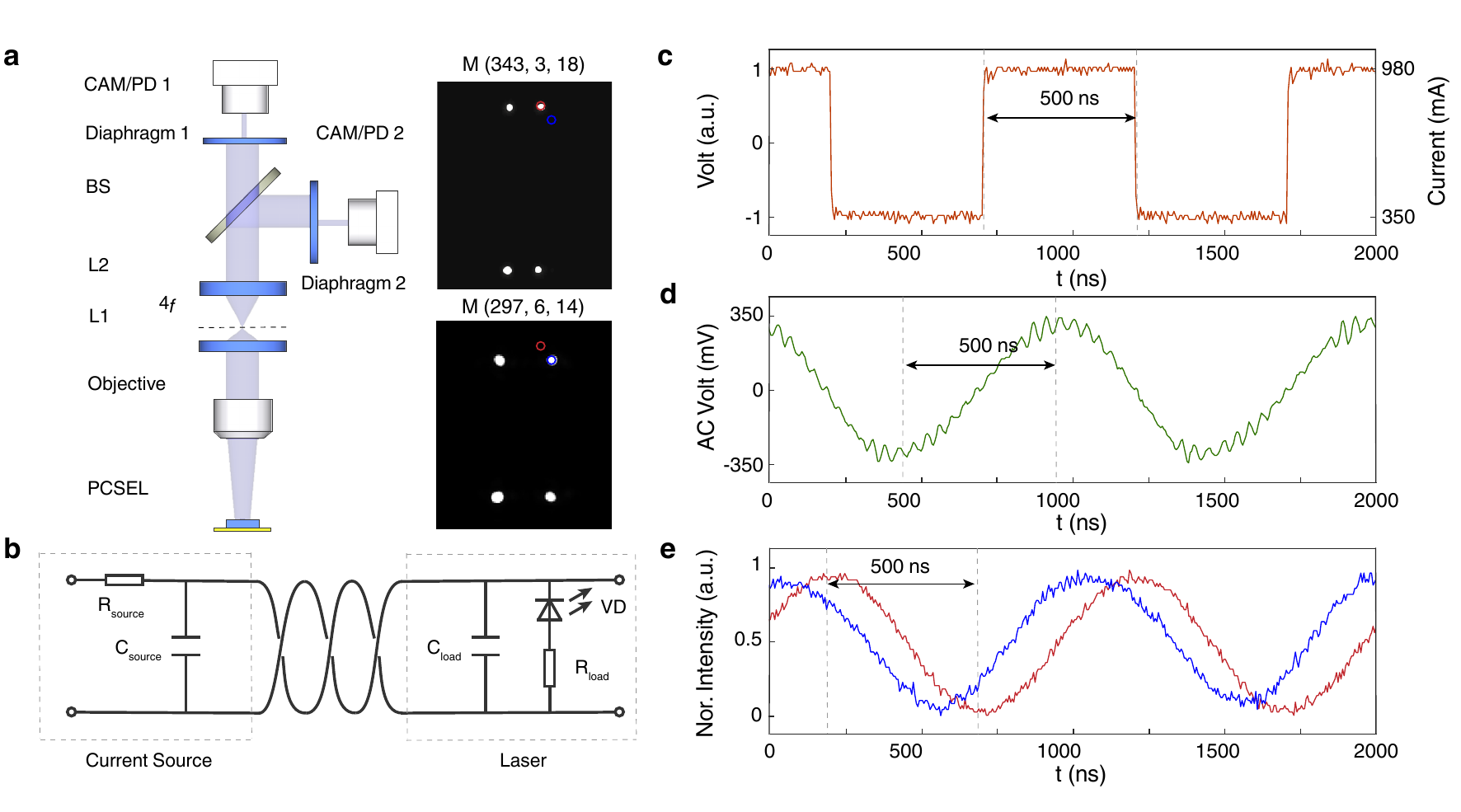}
\caption{ \textbf{Speed evaluation of active beam steering of the PCSEL}.
\label{mfig5}
(a) 
Schematic of the experimental setup to evaluate dynamic beam steering. L, Lens; BS, beam-splitter; CAM, camera; PD, photo-diode; Lens L$1$ and L$2$ are confocal. The observed far-field patterns of mode $(343,3,18)$ at $350$ mA and mode $(297,6,14)$ at $980$ mA. The red and blue circles mark the position of pin-holes in the far-field. 
(b) The circuit model represents that the laser is driven by a current source.
(c) The time-domain waveforms of current excitation, which periodically flip between two working currents $350$ mA and $980$ mA in a step duration time of $500$ ns.
(d) The AC voltage measured between the laser contacts. 
(e) The normalized responses of light intensities measured from the PDs, showing that the far-field patterns fast scan between mode  $(343,3,18)$ and mode $(297,6,14)$ in a transition time of $500$ ns.
}
\end{figure}

All the tests of PCSEL are carried out at room temperature. For the speed test, a bipolar waveform is applied to switch the dual level injection currents and the phenomena of far-field patterns scanning are observed by two PDs. The far-field patterns are shown in SFig.~6b and Fig.~\ref{mfig3}g, respectively. The PCSEL is driven by a current source at the rates of $400$ kHz, $600$ kHz, $800$ kHz, and $1$ MHz. The peak-to-peak values of $1100$ mV are measured from the oscilloscope. The measured time-domain waveforms of bipolar current excitation,  alternating current (AC) voltage, and the responses of light intensities under different speeds are shown in SFig.~7 and Fig.~\ref{mfig5}c-e. The beam steering runs at a maximum speed of $1$ MHz, due to the bandwidth limit of the current source.

\textbf{Conflict of interest}

The authors declare that they have no conflict of interest.

\textbf{Acknowledgments}
 This work was partly supported by the National Key Research and Development Program of China (2021YFB2801400), the National Natural Science Foundation of China (61922004, 62135001, and 62205328), National Key Research and Development Program of China (2020YFB1806405), the Major Key Project of PCL (PCL2021A14), and Huawei Technologies Co. Ltd. Grant TC20220323035 on the lasers.
 The simulation of this work was supported by High-performance Computing Platform of Peking University.

\textbf{Author contributions}

Mingjin Wang and Chao Peng conceived the idea. Mingjin Wang, Zihao Chen, Chao Peng, and Wanhua Zheng performed the theoretical study. Mingjin Wang, Zihao Chen, Yuanbo Xu, and Chao Peng performed the analytical calculations and numerical simulations. Mingjin Wang, Zihao Chen, Yuanbo Xu, Jingxuan Chen, Jiahao Si, and Zheng Zhang conducted the experiments and analyzed the data. Mingjin Wang, Zihao Chen, Chao Peng, and Wanhua Zheng wrote the manuscript, with input from all authors. Chao Peng and Wanhua Zheng supervised the research. All authors contributed to the discussions of the results.

\textbf{References}

\bibliography{Reference.bib}{}
\bibliographystyle{naturemag}

\clearpage

\clearpage

\clearpage

\clearpage

\clearpage

\end{document}